# Emergence of Griffiths phase, re-entrant cluster glass,metamagnetic transition and field induced unusual spin dynamics in $Tb_2CoMnO_6$

**Khyati Anand[1], Arkadeb Pal[1], Prajyoti Singh[1], Md. Alam[1], Amish G Joshi[2], Anita Mohan[1] and Sandip Chatterjee[1,#]**

[1]Indian Institute of Technology (BHU) Varanasi 221005, India

[2]CSIR- Central Glass & Ceramic Research Institute, Ahmadabad 382330, India

[#]Corresponding Email: schatterji.app@iitbhu.ac.in

## Abstract

The structural and magnetic properties of double perovskite$Tb_2CoMnO_6$ have been investigated. Electronic structure analysis by XPS study reveals the presence of mixed oxidation state ($Mn^{4+}/Mn^{3+}$ and $Co^{2+}/Co^{3+}$) of B-site ions. The dc and ac magnetization measurements reveal different interesting phases such as Griffith phase, re-entrant spin glass, metamagnetic steps, Hopkinson like peak and also unusual slow relaxation. The M-H curve indicates the presence of competing AFM/FM interactions. The disorder in $Tb_2CoMnO_6$ leads to spin frustration at low temperature givingrise to the re-entrant spin glass. Moreover, the field-dependent ac susceptibility studiesunraveled the presence of Hopkinson like peak associated with the domain wall motion and the large anisotropy field.The further study yielded that the relaxation associated with this peak is unusually slow.

## Introduction

The materials that possess both the magnetic and electric orders have received intense research attention globally owing to their potential for practical applications in next generation spintronic devices [1–5]. The coupling among the different order parameters allows one to have freedom of an additional gauge for monitoring one of these order parameters by the other which opens up unprecedented opportunities to achieve new functionalities in such materials [1–8]. Hence, an invigorated research attention has been bestowed on discovering such multifunctional materials and there is also an on-going search for the new mechanisms leading to such coupled order parameters.

Particular attention has been given to the magnetic oxides comprising the metal cations and the oxygen anions due to their abundant nature and high stability. Among such materials, the double perovskite (DP) compounds having the formula $A_2BB'O_6$ (A= Rare earth ions or alkaline ions; B/B'= transition metal ions) have received tremendous research attention world-wide for their wide span of interesting properties [2,7,9–17]. The physical properties of the DPs are profoundly influenced by its B-site structural ordering and its electronic structure [9,12,13,17,21–23]. The B/B′ ions having similar charge states and/or ionic radii triggers the anti-site disorder (ASD) in the system which is realized as the site-exchange among the B/B′ ions [13,17]. As a matter of fact, the ASD in DPs are known to intensely affect its physical properties especially its magnetic properties leading to the emergence of various exotic states viz., spin-glass states, Griffiths phase, exchange bias effects, and meta-magnetic transitions etc [7,12–15,17]. The ordered DPs exhibits ferromagnetism unlike its end members i.e. $ABO_3$ and $AB'O_3$ (B/B′=Mn,Co,Nietc) which are usually antiferromagnetic in nature [12,24]. The ferromagnetism in such ordered DPs are understood by the $180^0$ positive super-exchange interactions between B=$Co^{2+}$/$Ni^{2+}$ and B′=Mn following the Goodenough-Kanamori rule [24,25]. However, some $Co^{3+}$/$Ni^{3+}$/$Mn^{3+}$ ions creep into the system during the sample preparation which is unavoidable. Moreover, the similar charge states of these B/B′ ions also promote the ASD in the system [9,17,21]. As a consequence, additional anti-ferromagnetic (AFM) interactions come into play in the system via super-exchange interactions $Co^{3+}$—$O^{2-}$—$Co^{3+}$, $Co^{3+}$—$O^{2-}$—$Mn^{3+}$, $Mn^{3+}$—$O^{2-}$—$Mn^{3+}$, $Co^{2+}$—$O^{2-}$—$Co^{2+}$, $Mn^{4+}$—$O^{2-}$—$Mn^{4+}$ [12,15,16]. On lowering the temperature, a competition between the ferromagnetic (FM) and antiferromagnetic (AFM) interactions commences raising the spin frustration in the system which on reaching a critical level leaves the system in a randomly frozen state, known as the spin glass (SG) state [12,16,26–29].

Therefore, the SG states observed in such DPs are the manifestation of the nano-scale inhomogeneity which is the subject of the prime interest in solid state research in recent times owing to its complexity and interesting aspects viz., memory and aging effects, slow relaxation, thermo-magnetic irreversibility etc [28–35]. The underlying physics of such non-ergodic states still remained poorly understood and a complete understanding of such disordered magnetic states is an open challenge. In recent past, such glassy and non-equilibrium states have been observed in some systems where the basic entities responsible for such glassy spin dynamics are the bigger "spin clusters" instead of the single spins, such states are known as the "cluster glass" state [23,28,29,33]. On the other hand, due to existence of the quenched disorder, sometimes few systems enter in a glassy states at lower temperatures even after showing a long range magnetic ordering (LRO) at higher temperatures. Such lower temperature glassy states are known as the re-entrant spin glass/cluster glass states (RSG or RCG) [23,36,37]. Although there are ample of reports available on the systems exhibiting the spin-glass states, the reports on the RSG or RCG systems are particularly limited. Despite of the intense research interests on such systems to explore the true origin and nature of such re-entrant glassy states, it still remained controversial. Notwithstanding the complexity, RSG state was described reasonably well by mean-field model as used by Sherrington- Kirkpatrick for Ising spin systems and the model introduced by Gabay and Toulouse for Heisenberg spin systems [38,39]. According to this model, LRO parameter still remains in the RSG state, briefly which can be described as a state where both the spin-glass state and the long range magnetic correlation co-exist.

Apart from this, the quenched disorder is also known to be a key ingredient for the emergence of a peculiar magnetic phase where it neither behaves like a ferromagnet nor a paramagnet. This special phase is known as the Griffiths phase (GP) [14,19,23,40–46]. In this GP regime, the magnetization of the system fails to follow the typical Curie-Weiss law above the magnetic ordering temperature up to a certain critical temperature known as Griffiths temperature ($T_G$) above which it enters in a purely paramagnetic state. Although the experimental realization of such special phase after its theoretical anticipation was thought to be remote, the magnetic susceptibility study at different lower fields has provided a fine gauge for probing this GP [42]. As a matter of fact, the GP evolves in a system by the development of finite-sized clusters having short range ordering in the global paramagnetic matrix.

As compared to the most intricately studied La and Y based DP systems, the DP $Tb_2CoMnO_6$ is a less explored system and hence, exploring its properties may unravel new interesting magnetic states [47,48]. Moreover, the one end member of this DP system is $TbMnO_3$ which has drawn considerable research interest for last two decades owing to its magnetism driven ferroelectricity [49–51]. For last two decades, rigorous theoretical and experimental studies have been bestowed on this particular system for exploring its origin as well as to realize this coupled phenomenon at elevated temperatures. On the other hand, $TbCoO_3$ is a member of the $RCoO_3$ cobaltite family which has drawn considerable research attention globally for its thermally assisted spin-state transition of the $Co^{3+}$ ions. Hence, in the present work, we have chosen $Tb_2CoMnO_6$ for investigating its electronic structure as well as magnetic properties.

**Experimental Details**

The polycrystalline sample of $Tb_2CoMnO_6$ has been prepared by the conventional solid-state reaction method. Highly pure $Tb_4O_7$, CoO, and$Mn_2O_3$(>99.99%) oxide powders as precursors were taken in exact stoichiometric ratioto prepare the sample.The powder after having an hour of intimate grindingwasgiven the heat treatment at 1000°C for 24 hours in the air. This powder was re-ground and again was heatedat 1100°C for 36 hrs. This was followed by several heating cyclesat 1200°C with intermittent grindings for several days. This process was performed until we get a homogenous phase. The obtained powder in the last step was pelletized and sintered at 1300°C for 36 hours which was followed by the slow cooling rate to reduce the anti-site disorder.

**Characterizations**

Powder X-ray diffractogram (XRD) of the sample wasrecorded in a RigakuMiniflex II X-ray diffractometer. Rietveld refinement of the XRD patternwas done by FULLPROF suite software. The XPS experiment was carried out by an Omicron multi-probe surface science system with photon energy 1486.7 eV of monochromatic X-ray source Al-K line. The system has a hemispherical electron energy analyzer (EA 125). The average base pressure during the experiment was $5.6 \times 10^{-10}$torr.All the magnetization measurements (dc and ac) were performed by the superconducting quantum interference device (SQUID-VSM) based magnetic property measurement system (Quantum Design-MPMS).

**Result and Discussions**

**X-ray diffraction**

The Rietveld refinement of the XRD patternrecorded at room temperature (300 K) has been shown in Fig.1 which confirms that the sample has been prepared in single phase without any chemical and/or phase impurity. The well-refined pattern with $P2_1/n$ ($\chi^2$=2.58) space group suggests that the sample was crystallizedin a single phase monoclinic structure.All the crystallographic information regarding the bond lengths, bond angles and lattice constantsis summarized in table 1. The reduced bond angles (<$180^0$) of Mn-O-Mn and Co-O-Coindicate that significant octahedral distortion is present in theMn/$CoO_6$octahedra which is essentially triggered by the smaller ionic radius of the $Tb^{3+}$ ions. A quantitative measure of such distortion can be estimated by the formula: $\delta=(180°-\Phi)/2$ where $\Phi$ is the value of Mn-O-Mn/ Co-O-Co angles . In the present case, the value of $\delta$ is found to be~14.425 which suggests that the system has a significantly large distortion in $CoO_6$/$MnO_6$octahedra.

**X-ray Photoemission Spectroscopy Study**

The X-ray photoemission spectroscopy (XPS) is a versatile tool to probe the chemical valence states and the ligand coordination of a system. In Fig.2, the XPS survey scan ofthe system is presented. All the peaks are assigned by the national Institute of standard technology (NIST) confirming the presence of Tb, Co, Mn, O ions in the system. It also confirms that there are no impurities present in the system other than C. The presence of C is due to the absorption of it by the surface of the samplein the air exposure. The whole analysis has been done after making carbon correction with C1s line positioned at 284.8 eV to avoid the charging effect error. Fig. 3(a) depicts the core level XPS spectra ofTb 3d. It is divided into two major spin-orbit coupling peaks $3d_{5/2}$ and $3d_{3/2}$situated at 1241.3 eV and 1275.8 eV respectively. The line shape and the peak positions of the Tb3d spectra indicate the trivalent oxidation states for the Tb ions. Fig.3(b)showsthe XPS spectra of core level Mn 2p. The spectra are mainly composed of two main peaks of $2p_{3/2}$ and $2p_{1/2}$arising due to spin-orbit coupling. The two peaks are positioned at 653.5 eV and 641.7 eVwithdoublet separationof 11.7eV. In fact, the doublet separation for the Mn2p XPS spectra in $MnO_2$is reported to be ~11.8eV and for $Mn_2O_3$ is 11.6 eV. In our system, observation of the intermediate value of the doublet separation suggests that Mn ions are present in the mixed oxidation states. For further confirmation, we deconvolutedMn2p XPS spectra.As evident from the Fig. 3(b), the deconvolution analysis suggested for the presence of significant amount of $Mn^{3+}$ions along with the $Mn^{4+}$ ions.A shake up satellite peak of $Mn2p_{1/2}$ is clearly visible around 664.6 eV which agrees wellwith previously reported Mn 2p XPS spectra of a compound having Mn ions in mixed valence states [52].The further confirmation was sought through the study ofthe Mn 3s spectra (as shownin the inset of Fig. 3(b)) which has capability of probing the different charge states of Mn precisely. As a matter of fact, the Mn 3s doublet separation is arising due to the exchange splitting (ΔEex) of the parallel and anti-parallel coupling of a hole in 3d state and electron in a 3s state [53]. Thus, the separation of these two peaks is linearly related to spin of Mnions [53,54].It is wellestablished that exchange splitting of Mn 3s spectra is decreased with increased valency of Mnsuch as: for $Mn^{4+}$,it is 4.5 eV and 5.4 eV for $Mn^{3+}$ [53,55]. Hence for obtaining quantitative information about Mn valence state, Mn 3s spectra are important. $\Delta E_{ex}$ for the present system is found to be 4.8 eV and we can estimate itsMnvalency quantitatively by the linear relation between $\Delta E_{ex}$ and$\nu_{mn}$ [53,54]-

$$\nu_{mn} = 9.67 - 1.27\Delta Eex/eV$$

Where $\nu_{mn}$is the Mnvalency and $\Delta E_{ex}$ is exchange splitting in eV.For our system, the effective oxidation state of the Mn ions comes out to be ~3.57which is consistent with the Mn 2p spectra showing mixed valence states.

The XPS core level spectra of Co 2p are shown in Fig. 3(c). The study of core level Co 2p spectra is important as its satellite peaksare very sensitive towards the oxidation states of Co

ions.The main photoemission lines in Co 2p XPS spectra are associated with the well-screened states while its satellites are related to the poorly screened states. For CoO where the Co ions exist in divalent states, it shows strong satellite peaks while it is almost absent or relatively feeble in $Co_2O_3$ or $Co_3O_4$ respectively.Hence, the observation of the clear satellite peaks in the Co2p XPS spectra of the present system indicates the existence of $Co^{2+}$ ions significantly.Apart from this, the reported value of the doublet separation (DS) for CoO is 15.9 eV and $Co_3O_4$ is 15.3 eV [17,56]. For TCMO, DS of Co2p XPS is found to be ~15.5 eV suggesting the mixed valence states of the Co ions. For further confirmation, the peaks have been deconvoluted. The deconvolutionanalysis of the spectra also suggested that both the $Co^{3+}$ and $Co^{2+}$ions are present in the system.

The O1sXPS spectra are shown in Fig. 2(d) Two peaks are observed in the spectra where the first peak positioned at 529 eV is the characteristic feature of $O^{2-}$ ion of lattice oxygen [17] while the other peak at 530.6 eVis ascribed to the less electron rich oxygen species/absorbed oxygen species ( i.e$O_2^{2-}$ , $O_2^-$ or $O^-$).

## Magnetization study

In Fig.4(a), thetemperature (T) variation of zero field cooled (ZFC) and field cooled (FC) magnetization (M) curves with applied field 100Oehavebeen shown. Both the ZFC and FC curvesshowed an abrupt jump below 100K which suggests for the onset of the magnetic ordering in the system. Furthermore, to precisely identify the transition temperature,"dM/dT Vs T" is plotted (the inset of Fig. 4(b)). The inflection point of the curve is observed at $T_C$ ~99K which can be considered as the magnetic transition temperature. To further probe the nature of the magnetic transition, we have recorded the ac susceptibility ($\chi$) data. As evident from the Fig. 4(b), the ac $\chi'$ shows sharp and frequency independent peaks at ~99 K which is a clear indication of the onset of the long range magnetic ordering below this temperature.Apart from this, the dc ZFC-FC curves also showeda large bifurcation with irreversibility temperature $T_{irr}$~87K which can be attributed to the presence of magnetic spin frustration as well as the presence of strong magnetic anisotropy in the system [57,58]. In the case of large anisotropy present in the system, the ZFC curve shows a low value of magnetization while the FC curve starts increasing below the transition temperature and shows higher value of the magnetization (Fig. 4(a)) [59].In the dc magnetization curve, another anomaly is also observed below 7 K which can be presumably attributed to the ordering of rare earth element [14,50,60].For further investigations of the underlying physics, the ZFC curves have been recorded with the different applied field (Fig. 4(c)).It can be noted that for the lower applied magnetic fields, the curves showed a sharp cusp like peak which got broadened with increased magnetic fields.Moreover, the peak starts shifting towards lower temperature with increasing magnetic fields. This is a typical field dependence behavior observed in many other systems [58,61]. Although, this feature is observed in spin glass systems as well but it is not the sufficient condition for confirming the spin glass state. In fact, the observed behavior inthe present system may berelated to the local anisotropy field acting on the spin cluster [58]. The spins are frozen due to the competition between the local anisotropy and applied field which results in a cusp-like nature in $M_{ZFC}$. Further to confirm the nature of themagnetic transition, the isothermal field dependent magnetization (M-H) curveshave been recorded at 95 K and 300 K (Fig. 4(d)). The M(H) curve at 95 K shows nonlinear behavior just below the transition temperature,thus suggesting the onset of magnetic ordering.On the contrary, the M(H) curve recorded at 300 Kshows linear behavior suggesting the system is in purely paramagnetic state at this temperature.Furthermore, to estimate the effective paramagnetic moment ($\mu_{eff}$) and Curie-Weiss temperature ($\theta_{cw}$),the standard Curie Weiss (CW) law: $\chi^{-1} = \frac{H}{M} = \frac{3K_B}{\mu_{eff}^2}(T - \theta_{CW})$ was employed to fit the experimental "inverse susceptibility Vs temperature curve" in the paramagnetic region (inset of Fig. 4(a)). The fit yielded $\theta_{cw}$ ~ +34.59 K and $\mu_{eff}$~ 9.76 $\mu_B$.The positive value of$\theta_{cw}$..suggests towards the presence of the dominating ferromagnetic interactionsin the system. Furthermore, the large difference between the temperatures$\theta_{cw}$ and $T_C$indicates the spin frustration of the system. The theoretically calculated value of the spin only moment of the system TCMOis found to be 10.65 $\mu_B$(considering all the spins in high spin state) which is closely matching with the experimentally obtained effective paramagnetic moment..

Interestingly, the temperature variation of the inverse susceptibility study at different magnetic fields showed a peculiar down-turn deviation from the expected linear behavior of the paramagnetic state (i.e. violating the Curie-Weiss law) at temperatures well above the magnetic transition temperature $T_c$~99 K (Fig. 5). This particular down-turn feature is typically found in the systems showing the Griffiths phase (GP) [14,23,40,41,44].However, on increasing magnetic field, the down turn behaviour is observed to get softened and with sufficiently high field, it yields CW like behaviour, which is also a characteristic for GP [Fig. 5]. The Griffiths phase temperature is identified to be $T_G$ ~130 K below which the typical down-turn behavior of the inverse susceptibility curves start commencing [43,46,62].

In the GP regime, the susceptibility curves usually follow the power law

$$\chi^{-1} = (T-T_C^R)^{1-\lambda} \quad (1)$$

where λ ( 0< λ<1 ) isthe power law exponent which is a measure of the deviation from the ferromagnet [40,41]. In equation (1), λ ~ 0 refers to theCurie-Weiss law in GP regime and $T_C^R$is the critical temperature of the randomly diluted perfectly the paramagnetic state following the normal Curie-Weiss model.Usually, taking $T_C^R$=$\theta_{cw}$ is a good choice, since it ensures λ~0 in the paramagnetic region. However,systems in which the magnetic ordering temperature ($T_c$) and Curie-Weiss temperature ($\theta_{cw}$)have a large difference i.e. $\theta_{cw}$<<$T_c$; putting $T_C^R$=$\theta_{cw}$ gives errorful estimation of λ. In the present system, there is also a relatively large difference between $T_c$ and $\theta_{cw}$due to the presence of frustration of the system. In such systems, a reasonable choice is to choose $T_C^R$=$T_c$ which will solve this confusion since the power law will be fitted in Griffith phase region above the transition temperature [46]. The Fig. 6(a) is demonstrating the "$\log_{10}(1/\chi)$ Vs $\log_{10}(T- T_C^R)$" plotand its power law fitting in the linear region of Griffith phase above the transition temperature and well below the paramagnetic linear region. The fitting yielded λ~0.87 which clearly suggests the presence of GP in the system. Moreover, as evident from the inset of Fig. 6(a), theimaginary part of ac susceptibility curvesχ"(T) at different frequencies also showed an anomalyin the GP region [63]. The interesting feature which can be noted is the observation of the negative value of χ"(T).In fact, temperature-dependent imaginary part of ac susceptibility should be a positive quantity since it is the measure of energy dissipation. However, the observation of such negative χ"(T) is not new and was frequently reported in many systems previously [64,65].The χ"(T) curves are showing a clear down turn behavior in the GP regime which gets suppressed with lowering the frequencies. One possible reason of such negative value of χ"(T) could be the calibration error of the instrument. Hence, to investigate whether it is really related to the instrumental errors or it is intrinsic to the system,different measurementshave been performed which yielded similar result of the negative value of magnetization in a particular temperature region (Fig. 6(b),6(c) and 6(d)). Thus, it ruled out the experimental errors and suggested towards the intrinsic property of the system which is seemingly related to the Griffiths phase.Again, the spin relaxation in the GP regime is expected to be slower (owing to its short range correlations) than that in the paramagnetic region. Hence,toget further insights into the spin dynamics in the Griffiths like phase, the isothermal remanent magnetization(IRM) measurement is carried out with time.The spin dynamics of the diluted magnet in the GP regionshould follow either of the models as given below where C(t) is theauto-correlation functions of the spins [66]

$$C(t) \propto \exp(-A\left(lnt^{\frac{d}{d-1}}\right)): \text{ For Ising}$$

$$C(t) \propto \exp(-Bt^{1/2}): \qquad \text{For Heisenberg System}$$

To examine whether the spin relaxation in the GP region of the present system follows such prediction or not, we have recorded the IRM data at two different temperatures in the GP regime (Fig. 6(c) and 6(d)).The sample is cooled from 300K to the desired IRM temperatures (110 K and 125 K) with the presence of magnetic field H= 500 Oe.Thereafter, the measurement the decay of the magnetization as a function of time was recordedafter switching off the magnetic field. Our IRM study shows that the observed decay of the remanent magnetization follows spin auto-correlation function for the Heisenberg model which is essentially expected for the GP region. The fitting yieldslack of agreement with the exponential model, therefore, ruling out the pure PM phase above $T_c$.It is again interesting to note here that the remanent magnetization is found to be negative after the immediate removal of the magnetic field (which was positive) and with the evolution of the time, the magnetization is relaxing towards zero. We have repeated the experiment several times so as to confirm the reproducibility of the data and we found that every time the remanent magnetization becomes negative at t=0 s. This might be associated to the inherent anisotropy of the system which is dominating since the applied cooling magnetic field was low. We can reiterate that the $\chi''(T)$ curve also showed negative value.To further re-investigate the occurrence of the negative magnetization,wehave carried out the thermo-remanent magnetization (TRM) measurement (Fig. 6(b)). As a matter of fact, our TRM measurement also supportedthe previous results of the IRM data by exhibiting a drastic dip leading to the negative magnetization at temperaturesjust above the transition temperature (99 K). The plausible reason might be the domain wall motiondue to the complex magnetic domain wall structure of the present system [65]. To further elucidate the GP, the role of the quenched disorder in the form of the ASD may be considered which is known to play a significant role in hindering the long range ordering and thus leading to the formation of the short range correlated clusters [67]. Hence, the presence of the ASD in the system which is evident by different analysis as mentioned above may play a significant role in bringing out the GP in the present system by giving rise to random exchange interactions. Additionally,in this system, $Co^{2+}$ and $Mn^{3+}$ ions are Jahn-Teller (JT) active ions which result in the formation of different bond lengths due to crystal field splitting in $e_g$ leveland leaves the system frustrated [63]. Thus, theJT distortion creates static quenched disorder which may also contribute to the observed GP phase [68].

Apart from this, the presence of the anti-site disorder isalso suggested by the observation of the reduced saturation magnetization ($M_s$) at 5Kwhich is displayed in the Fig. 7(b). The $M_s$is estimated by the linear extrapolation method which comes out to be 14.98$\mu_B$/f.u. On the other hand, the theoretical saturation magnetic moment can be calculated by the formula [2gJ +6]$\mu_B$, where 6$\mu_B$arises from the FM alignment of the Co/Mnsublattice and gJ is related to the rare earth moment [69,70]. The value of the theoretical moment is calculatedto be 24$\mu_B$/f.u which is larger than experimental $M_S$. This deviation could be produced by the partial antiferromagnetic alignment Co/Mn moments (due to the presence of the ASD and $Co^{3+}$/$Mn^{3+}$ ions) and canting of rare earth element at low temperature. Hence, the antiphase boundaries originated by the inherent antisite disorderof the system can also contribute to the reduction of the observed moment in addition to compensation of magnetic moment as required by the magnetic structure. The M-H loops at different temperaturesare shown in Fig. 7. The M-Hcurves observed at relatively higher temperatures (25 K and 50 K) show characteristically different behavior than those observed at lower temperatures (T= 2 K & 5 K). However, the hysteresis curves observed at 2 K and 5 K show almost similar patterns while the only noticeable difference that can be observed is the slightly lower magnetization in lower temperature M-H curve (inset of Fig. 7(a).This result may be associated with the well-established antiparallel alignment of paramagnetic $Tb^{3+}$ spins by the internal field of the FM Co/Mnsublatticeat low temperature [14]. Our M(T) (ZFC curve) also supports this result of the onset of the Tb ordering by showing an anomalybelow 7 K. Another possible reason for smaller magnetization at low temperature (below 5 K) might be due to the canting of spin. As the temperature decreases, the canting between the spins increases and gives rise to the lower value of magnetization [16]. However, the hysteresis curvesat both the temperaturesare not showing saturation up to the maximum applied field 5 T.As already stated, the ASDs lead todifferent antiferromagnetic couplings which give rise tosuch unsaturated M-H loops.However, the large hysteresis is indicating the predominant ferromagnetic interactions presentin the system.Thus there is a significant competition of FM and AFM interactionsin the present system.

On the other hand, in the M-H loop recorded at 50 K showed a sudden slope change after reaching a certain critical field forming a step like feature which is a featureof themetamagnetic transition of magnetic materials (Fig. 7(d)). Thismetamagnetic feature is less prominent at 25 K and disappearson further lowering the temperature (Fig. 7(c)).It is relevant to mention here that meta-magnetic steps have been previously observed in DP compounds $Y_2CoMnO_6$, $Eu_2CoMnO_6$ and $Lu_2CoMnO_6$ where the A site is occupied by the non-magnetic ions [13,71,72].The plausible originof the observation of such meta-magnetic stepsmay be related to the ASDs which may occur as a point defect or in the group to form antiphase boundaries (APBs) producing as planar defects [13]. In Co/Mn based DP systems, the existence of the APBsis commonly observed and strongly pinned magnetic domain walls are generated by the interaction between themagnetic ions at the APB-domain boarders [13]. The possible underlying mechanism for the APB formation in the present system could be such thattheyare produced by the same number of regions which are enriched in $Co^{2+}$or $Mn^{4+}$ ions.Due to the existence of the strong AFM interactions across the APBs via $Co^{2+}$— $O^{2-}$—$Co^{2+}$/ $Mn^{4+}$—$O^{2-}$—$Mn^{4+}$ , the neighboring FM domains formed due to the $Co^{2+}$—$O^{2-}$—$Mn^{4+}$ super-exchange interactions are forced to align anti-parallel to each other. Consequently, the saturation magnetization of the M(H) loops gets diminished as well as the strong anisotropic pinning forces across the APBs of the anti-parallel FM domains offer strong hindrance to the applied magnetic field to orient them along the field direction. Thus, on reaching a critical value of the magnetic field, sudden rise in the magnetization in the form of the steps is observed. Different size and nature of APBs result in the occurrence of steps at different field in the M-H loop. In contrast to the $Y_2CoMnO_6$ and $Lu_2CoMnO_6$ systems which are reported to show pronounced steps in their M-H loops down to the lowest temperatures (2 K) [13,71], the present system $Tb_2CoMnO_6$exhibited meta-magnetic steps only at higher temperatures and on lowering the temperature, the steps disappeared. This is seemingly associated to the 3d-4f interactions owing to the magnetic nature of the Tb ions.

The presence of the spin frustration in the present system was evident from different studies as already discussed. Moreover, observation of an anomaly below 40 K in the M(T) curves may indicate the emergence of a secondary phase at this temperature(Fig. 4(a)). Unlike the dc susceptibility measurements, ac susceptibility measurement is a magnificent tool to probe into the spin dynamics and makes it possible to investigate glassy behavior [26].Hence to probe the spin dynamics of the present system, we have studied its ac susceptibility measurements at low temperature range (<50 K). Fig. 8(a) is showing temperature variation of $\chi_{ac}''$. The clear peaks in $\chi_{ac}''(T)$ are observed at ~ 33 K. The observed broad and clear frequency dependent $\chi_{ac}''(T)$ peaks are suggesting that the system enters in a glassy state below 40 K. Such peaks are the manifestation of the underlying slow spin relaxations in the glassy state [23,26,29]. On the contrary, the long-range ordering peaks are typically very sharp (λ-like) in natureunlike the observed broad glassy peaks [12,26]. Hence, the observation of such lower temperature glassy peaks despite having a higher temperature long range ordering is a typical feature of a re-entrant spin glass (RSG) state. The intriguing physics behind such RSG state can be understood based on the presence of both competing FM-AFM interactions in the system. On lowering the temperature, a peculiar magnetic state is achieved at which the strength of both FM and AFM interactions becomealmost equaland leaves the spins in the frustrated situation leading to the glassy behavior at lower temperatures [73,74].However, if one of these competing interactions (FM or AFM) is relatively stronger than the other, this will result to produce a cluster glass (CG)state where the spin frustration or disorder exists locally in the form of small clusters in the CG state [29,74]. In such states, the basic entity responsible for the slower relaxation is not simply the individual spins rather they are the clusters.

For getting further insights into RSG state, we have fitted the data in different models. From the frequency dependence of the ac peaks, we have calculated Mydosh parameter (p) which is a universal tool to distinguish SG state [26]. Here,

$$\boldsymbol{p} = \frac{\Delta \boldsymbol{T_f}}{\boldsymbol{T_f} \Delta \boldsymbol{log_{10}(f)}},$$

Where $\Delta T_f = T_{f1} - T_{f2}$ and $\Delta log_{10}(f)$=$log_{10}(f1)$-$log_{10}(f2)$. For, typical SG or CG systems, p lies between 0.005 and 0.08, while for the superparamagnetic system, it is greater than 0.2. The obtained value of p ~ 0.07 for TCMO confirms the presence of glassy state.

Moreover, in an SG or CG state, the spin dynamics gets slowed down below the critical temperatures $T_f$. This critical slowing down of spins near $T_f$ can be investigated using the dynamic scaling law [23,29]

$$\tau = \tau_0 \left(\frac{T_f - T_G}{T_G}\right)^{-z\nu}$$

Where $f$ is the excitation frequency corresponding to the characteristic spin-flip time ($f_0$=$\frac{1}{\tau_0}$); $T_G$ is the equivalent spin glass freezing temperature in the limit of $f \rightarrow 0\ Hz$ *and* $H_{DC} \rightarrow 0\ Oe$, $f_0$ is related to the characteristic spin flipping time ($\tau_0$) as $f_0$=$\frac{1}{\tau_0}$; zv is the dynamical critical exponent. In Fig. 8(b), "$\log_{10}$Vs $\log_{10}(\frac{T_f - T_G}{T_G})$" curve has been plotted and the best fitting with the above dynamical scaling law yielded: $f_0$~$3.5\times10^5$Hz ($\tau_0 = 2.82\times10^{-6}$s), $T_G$ = 27 K (which is near to the observed glass freezing temperatures) and the exponent zv is found to be ~ 4.4 which is satisfactory for glassy state (4<zv< 12). For a canonical SG system, $\tau_0$ typically lies between ~$10^{-12}$-$10^{-13}$ s which is less than the observed value ~$10^{-6}$s by few orders. The larger spin flipping time is suggesting the observed transition is due to freezing of finite-sized clusters (which take more time to relax) rather than individual spins [23,28,29].

For further investigations of inter-cluster interactions, the empirical Vogel-Fulcher (VF) law can be employed to fit the above curve "$f$ vs$T_f$". The law being of the form [29,75]

$$\tau = \tau_0 \exp(-\frac{E_A}{K_B(T_f - T_0)});$$

Where$f_0$ is a characteristic spin-flip time, $T_0$ is formally known as VF parameter which is a temperature representing the strength of inter-cluster interaction strength and $E_A$ is the activation energy.Fig. 8(c) shows the linear fitted graph "ln(τ) Vs $1/(T_f-T_0)$" using the V-F law. The best fitting yielded $\tau_0$ ~ $10^{-6}$Hz (which is of the same order of $\tau_0$ obtained from previous dynamic scale fitting), $T_0$=25.55 K and $E_A/K_B$=37.47 K. The comparable values of $T_0$ and activation energy indicate the existence of intermediate inter-cluster couplings in the system. Since the large value of activation energy than $T_0$ denotes weak coupling and smaller one indicates strong coupling [76]. The obtained large value of $\tau_0 = \frac{1}{f_0}$ is again suggesting the presence of interacting magnetic spin clusters.

Another experimental realization of slow spin relaxation in the SG or CG state can be found in the "time (t) evolution of remanent magnetization m(t) (TRM)" below $T_f$. The measurement was carried out following the field cooled (FC) protocol. The sample was cooled with a field H=0.1 T down to 25 K (below $T_f$) and the TRM data was recorded after switching off the magnetic field. The normalized magnetization m(t)=$\left(\frac{M_t}{M_{t=0}}\right)$ has been plotted as a function of time and shown in Fig. 8(d). The TRM data can be analyzed using KWW (Kohlrausch Williams Watt) stretched exponential equation as given below [31,32]:

$$m(t) = m_0 - m_g exp\left\{-\left(\frac{t}{\tau}\right)^{\beta}\right\};$$

Here, $m_0$ is associated to the initial remanent magnetization, $m_g$ is representing the magnetization of glassy component, $\tau$ is the characteristic relaxation time constant and $\beta$ is the shape parameter or stretching exponent. Another power law also often used for the analysis of TRM data is m(t) $\propto t^{\pm\alpha}$[72]. However, we tried to fit our TRM data with both the above relations but found that the best fitting is obtained with the KWW model, shown in Fig. 8(d). The fitting was not satisfactory for the power law ($t^{\pm\alpha}$), and not shown here. The KWW fitting is a powerful technique which is widely used for the investigations of the m(t) data for glassy or disordered systems [32]. For the different class of disordered systems, the $\beta$ value lies between 0 and 1. The obtained $\beta$ value for TCMO is ~ 0.44, thus confirming the existence of a glassy state at this temperature (25 K). Therefore, all the above facts confirm the system entering the RCG state at low temperatures. However, co-existence of high-temperature long-range ordering and low-temperature glassy state have been reported in systems such as double perovskite disordered ferromagnet $La_2NiMnO_6$, $La_{1.5}Sr_{0.5}CoMnO_6$, spiral magnet $BiMnFe_2O_6$, and antiferromagnet $Pr_2CoFeO_6$etc [12,16,23,31].Present system TCMO contains the major microscopic ingredient for glass transitions is B-site disorder [12,15,16,23,37]. The domain formation involves microscopic time scalesin pure FM or AFM systems but because of the presence of disorder, it attributes pinning of the domain wall which essentially causes to metastable states [37,77]. Thus in the experimental time scale system cannot achieve an equilibrium statewhich eventually caused a non-equilibrium phase like spin-relaxations, aging effects, etc [37,77].The B-site disorder gives rise to the local environment of the magnetic spins to be inhomogeneous in TCMO whichcreates random exchange bonds in the system.Thus, this spin frustration at low temperatures ends up in non-collinear, frozen, random states of spins leading to RCG state. Hence for TCMO, the high temperature ($T_C$ ~ 99K) long-range ordered FM state gets frustrated due to the increasing competition of AFM and FM interactions with lowering temperature, thus reaching to an RCG state. Eventually, an RCG state evolves when one of the competing interactions dominates the other, unlike the RSG state where both FM/AFM states are of equal order [29,58]. Thus, the dominating FM interaction in TCMO is presumably associated to the observation of the RCG state.

To further get the insight of different magnetic states and field effect spin dynamics, we have measured real and imaginary part of ac susceptibility ($\chi'$and$\chi''$) with different applied dc field. Fig. 9 and its inset are demonstrating $\chi'$ and $\chi''$ with frequency at different dc fields respectively. The unusual peak in $\chi'$is observed which becomes broader with increasing the applied field and is split into two peakswith further increase of the field. The peak below $T_C$shifts towards lower temperature side with increasing DC bias. On the other hand, the peak at a higher temperature (above $T_C$) shifts to the higher temperature side with the applied field. In the ferromagnetic state, the in-phase component $\chi'$ is depended upon two parameters as $\chi' \propto \frac{M_S(T)^2}{K(T)}$, where $M_S$ is the saturation magnetization at particular temperature T and K(T) is anisotropy energy density [61,78]. Therefore, the decrease in $\chi'$ below $T_C$ is due to the increase in anisotropy energy density K. Thus due to the large anisotropy energy, an additional peak has been observed below $T_C$because it blocks the spin and does not allow to respond in the magnetic field. This peak is named as Hopkinson like peak [43,61,79]. This very fast increase in anisotropy energy below the transition temperature is the result of continuous change in the size and shape of the FM cluster (domain wall motion). Similarly, in the inset of Fig. 9, the peak below transition is shifting in lower temperature side with field and peak at transition or just above the transition suppresses with the action of higher field. Similar to the typical FM-PM transition, in the present case also the peak shifts toward higher temperature with field and is suppressed in amplitude with the field (inset of Fig. 9), indicating clear magnetic transition [80].To further get the origin of this Hopkinson like peak we performed "$\chi'$ Vs temperature" with different frequencies at 750 Oe (Fig. 10).We found a very interesting feature in the peak lying immediately below the magnetic transition as it behavesdifferently forlower and higher frequencies. The peak even at very low frequency (at 3 Hz) is growing progressively in its amplitude which essentially indicates towards the unusually slow spin relaxation.This cannot be thermally activated relaxation of single spin since less frequency dependence has been observed at higher frequencies. The peak is shifting to a higher temperature with increasing the frequency which is quite similar to the spin glass state. But suppression in the amplitude of the peak and almost negligible frequency dependence at comparatively high frequencies show strong contrast with conventional spin glass phase. Therefore it rules out the possibility of a spin glass state. The similar observationof unusually slow relaxation has also been observed earlier in other systems [81,82]. The underlying physics of such slow relaxation was explained through the existence of oppositely spin-polarized regions arising due to the strong dipolar interactions in such systems [81,82].Hence, it is plausible to elucidate the observed similar unusually slow relaxation in the present system by the existence of some oppositely aligned giant domains triggered by the presence of the inherent anti-site disorder of this system. Thus it takes larger time scale due to the presence of larger thermal energy barrier while rotating in the direction of the applied field and consequentlyexhibits slow relaxation showing growth of the associated peak (in $\chi'$) even at 3 Hz. It is pertinent here to reiterate that the earlier discussions in the present system disclose the presence of antisite disorder and anti-phase boundary which result in an exhibition of different features such as metamagnetic steps in M(H), re-entrant spin glass phase and Griffith phase. Thus the aforementioned origin of the observed slow spin relaxation seems to be plausible for this system. However further theoretical and experimental studies may be helpful to get more insights into the underlying physics of observed phenomena.

## Conclusion

We have synthesized $Tb_2CoMnO_6$ polycrystalline double perovskite via conventional solid state reaction. Electronic structure analysis by XPS study reveals the presence of mixed oxidation state ($Mn^{4+}/Mn^{3+}$ and $Co^{2+}/Co^{3+}$) of B-site ions. The Mn 3s spectra indicate thatMn ions to be in $Mn^{4+}/Mn^{3+}$states The core level spectra of Mn 2p and Co 2p also support the mixed valence statesofMn and Co.The main aspect of the present work is the dc and ac magnetization studies which reveal different interesting phases such as Griffith phase, re-entrant spin glass, metamagnetic steps, Hopkinson like peak and also unusual slow relaxation in $Tb_2CoMnO_6$. The inverse of dc susceptibility shows downturn behavior at low fields which suppresses with increase of applied magnetic field. This is the prominent feature of Griffith like phase which has been further confirmed by the power law. The presence of inherent anti-site disorder along with mixed valence states of B-site ions and J-T active ions are the most important ingredientsfor the evolution of this interesting phase. The M-H curve is not saturated up to 5 T and the extrapolated saturation value is lower than that of the theoretically calculated. This might be the result of the presence of competing AFM/FM interactions. M-H curve at 25 K and 50 K also shows metamagnetic step which can be attributed to the drastic reorientation of the pinned domain. These domains are aligned antiparallel by the APBs at zero field. The disorder decreases the magnetic ordering, as well as the homogeneity of APBs giving rise to sudden slope changein the hysteresis loop. This disorder further leads to re-entrant spin glass. Analysis by different models yielded that the system entered in a glassy state below ~ 33 K. Moreover, the field-dependent ac susceptibility studiesunraveled the presence of Hopkinson like peak associated with the domain wall motion and the large anisotropy field. Further study yielded that the relaxation associated with this peak is unusually slow. Thus the present system exhibits different magnetic phenomena which are mainly associated with the presence of inherent antisite disorder.

Acknowledgment

The authors express sincere gratitude to the Central Instrumentation Facility Centre, Indian Institute of Technology (BHU) for their supports for the magnetic measurements facilities (MPMS).

**Table 1**: Structural parameters and crystallographic sites determined from Rietveld

| Temperature | 300 K |
|---|---|
| β | 90.067 |
| a (Å) | 5.28093 |
| b (Å) | 5.58929 |
| c (Å) | 7.51605 |
| V (Å$^3$) | 221.8484 |
| Tb | 4e |
| x | 0.51291 |
| y | 0.56778 |
| z | 0.24788 |
| $B_{iso}$ (Å$^2$) | 0.49980 |
| Co | 2d |
| x | 0.50000 |
| y | 0.00000 |
| z | 0.00000 |
| $B_{iso}$ (Å$^2$) | 0.49980 |
| Mn | 2c |
| x | 0.00000 |
| y | 0.50000 |
| z | 0.00000 |
| $B_{iso}$ (Å$^2$) | 0.49980 |
| O1 | 4e |
| x | 0.38180 |
| y | 0.95957 |
| z | 0.23377 |
| $B_{iso}$ (Å$^2$) | 0.49980 |
| O2 | 4e |
| x | -0.10370 |
| y | 0.65364 |
| z | 0.28356 |
| $B_{iso}$ (Å$^2$) | 0.75956 |
| O3 | 4e |
| x | 0.33596 |
| y | 0.69744 |
| z | 0.54637 |
| $B_{iso}$ (Å$^2$) | 0.49980 |
| $d_{Co-O(1)}$(Å) | 1.78547 |
| $d_{Mn-O(1)}$ (Å) | 2.15118 |
| $d_{Co-O(2)}$(Å) | 1.39060 |
| $d_{Mn-O(2)}$ (Å) | 2.83660 |
| $d_{Co-O(3)}$(Å) | 1.93753 |
| $d_{Mn-O(3)}$ (Å) | 2.00838 |
| <(Mn)-(O1)-(Mn)>(deg) | 145.11 |
| <(Co)-(O1)-(Co)>(deg) | 144.58 |
| <(Mn)-(O2)-(Mn)>(deg) | 151.15 |
| <(Co)-(O2)-(Co)>(deg) | 138.99 |
| <(Mn)-(O3)-(Mn)>(deg) | 127.37 |
| <(Co)-(O3)-(Co)>(deg) | 128.57 |

Figure Caption

Fig. 1: Reitveld refinement of XRD pattern collected at room temperature (300 K).

Fig. 2: XPS survey scan of TCMO sample at 300 K.

Fig. 3:(a) depicts the core level XPS spectra of Tb 3d, (b),(c) and (d) depict the core level XPS spectra of Mn 2p, Co 2p and O 1srespectively.

Fig. 4: (a) M(T) ZFC-FC curve at 100 Oe for TCMO. Inset presents the "Curie-Weiss fit to the $1/\chi$ vs T" plot. (b) Shows ac $\chi'$(T) curves at different frequencies. Inset of (b) shows "dM/dTvs T" plot. (c) Depicts" ZFCM(T)"curves at different fields. (d) Demonstrates M(H) curves at 95 K and 300 K.

Fig. 5:Shows the inverse susceptibility Vs temperature curves at different fields while its inset shows a closer view of the down-turn behavior.

Fig. 6: (a) Demonstrates the power law fitting to the log-log plot of "$1/\chi$Vs $((T-T_C^R)/T_C^R)$" and its inset shows "$\chi''$VsT" curves in the Griffith phase region. (b) Depicts the residual magnetization (M)Vs T curve (TRM) while (c) and (d) show IRM study with its Heisenberg, Ising and exponential model fitting at 110 K and 125 K respectively.

Fig 7: (a) describes M(H) curve at 2 Kwhile its inset shows enlarged view of of M(H) curves at 2 K and 5 K. (b), (c) and (d) show M(H) curves at 5 K, 25 K and 50 K respectively.

Fig 8: (a) presents $\chi''$(T) curves at different fields in the CG region. (b)Shows the dynamic scaling fit to the log-log plot of "$\tau(=1/f)$ Vs $(T_f-T_G)/T_G$)" and (c) shows the V-F fit to the log-log plot of "$\tau$ Vs $1/(T_f-T_0)$" . (d) KWW fit to the m(t) data recorded at 25 K.

Fig 9: Presents "$\chi'$Vs T" curves at different fields and its inset shows the corresponding"$\chi''$VsT"curvesat different fields.

Fig 10: Shows "$\chi'$Vs T" curves at different frequencies recorded with applied dc bias field of 750 Oe.

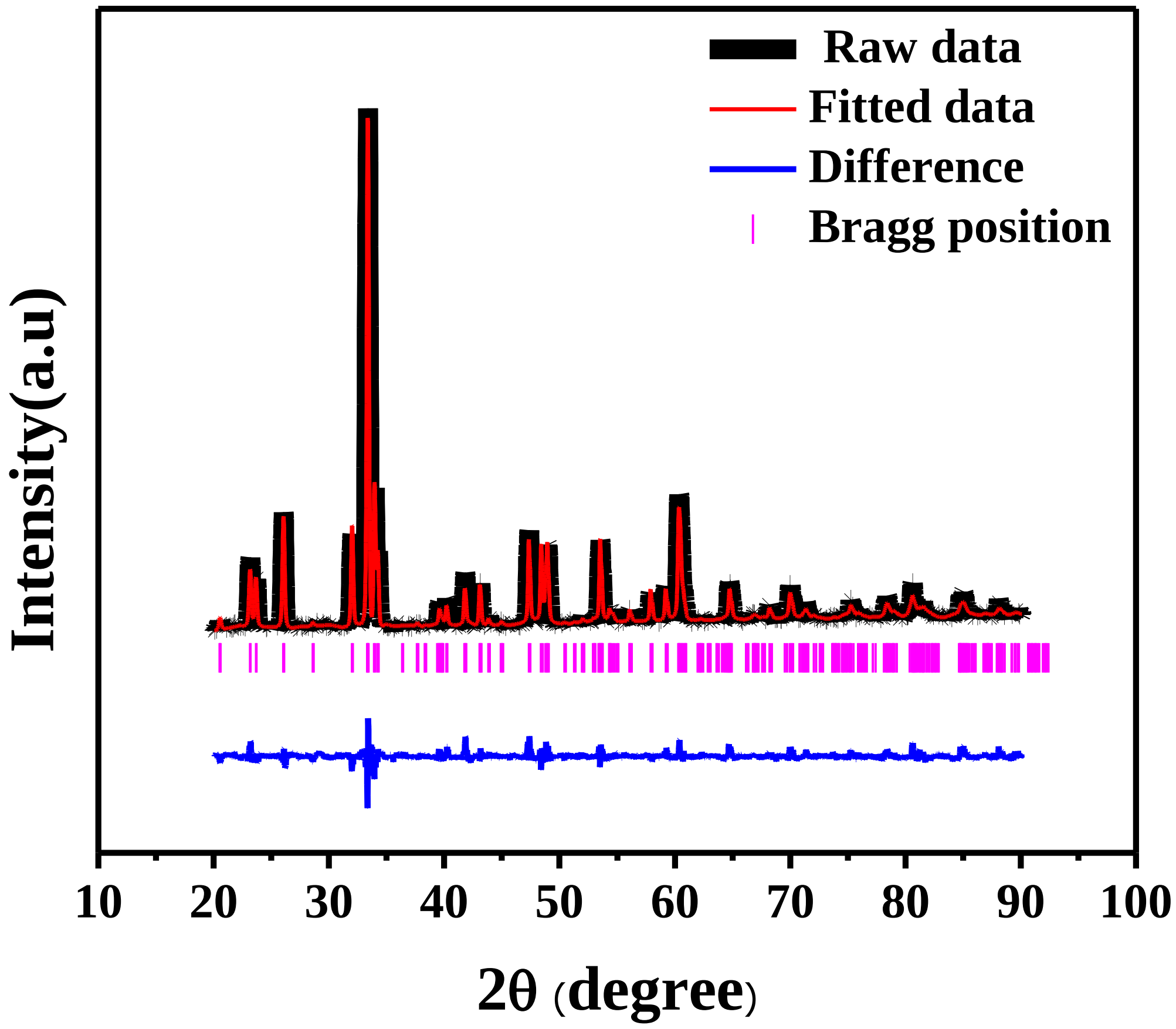

**Figure 1**

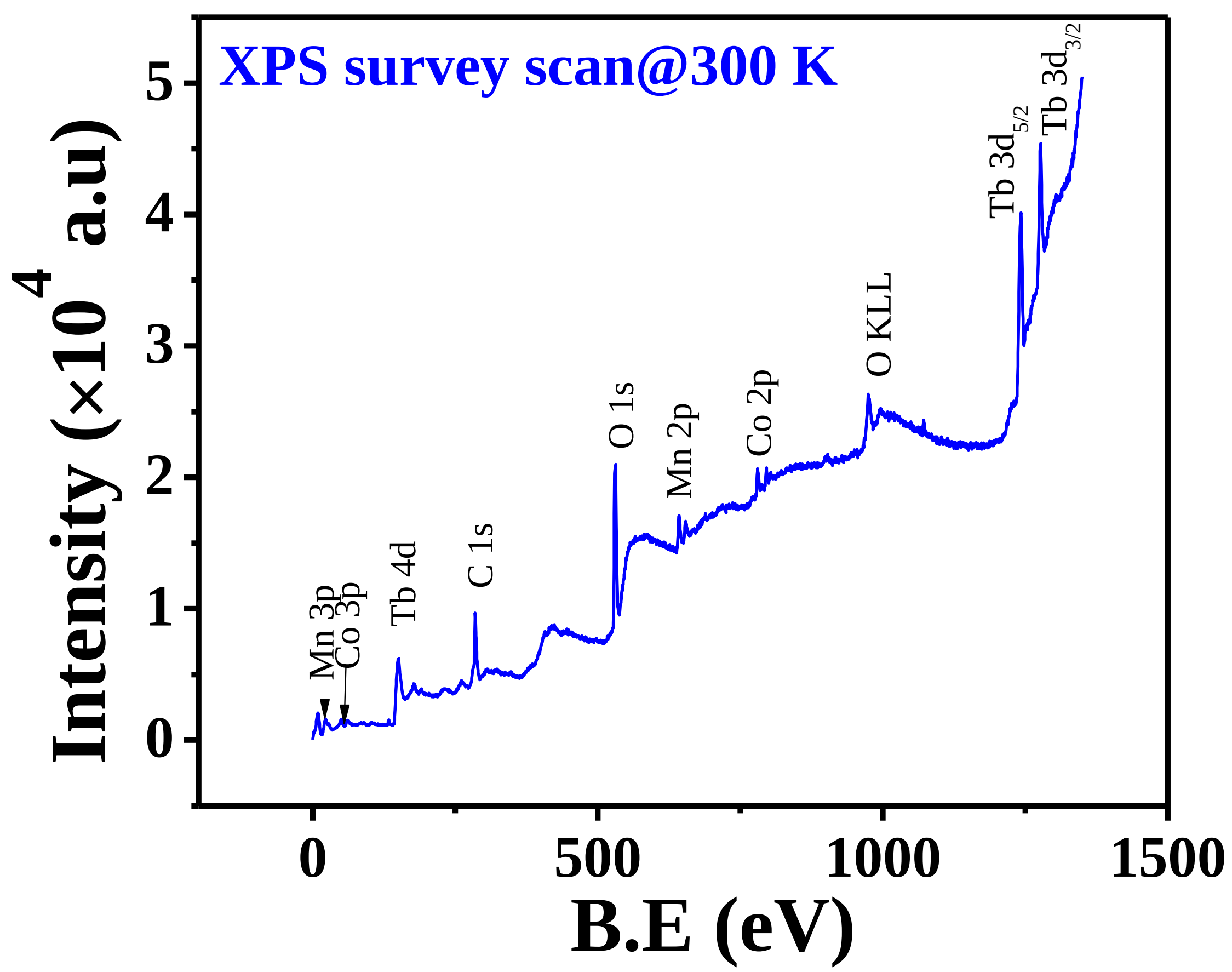

**Figure 2**

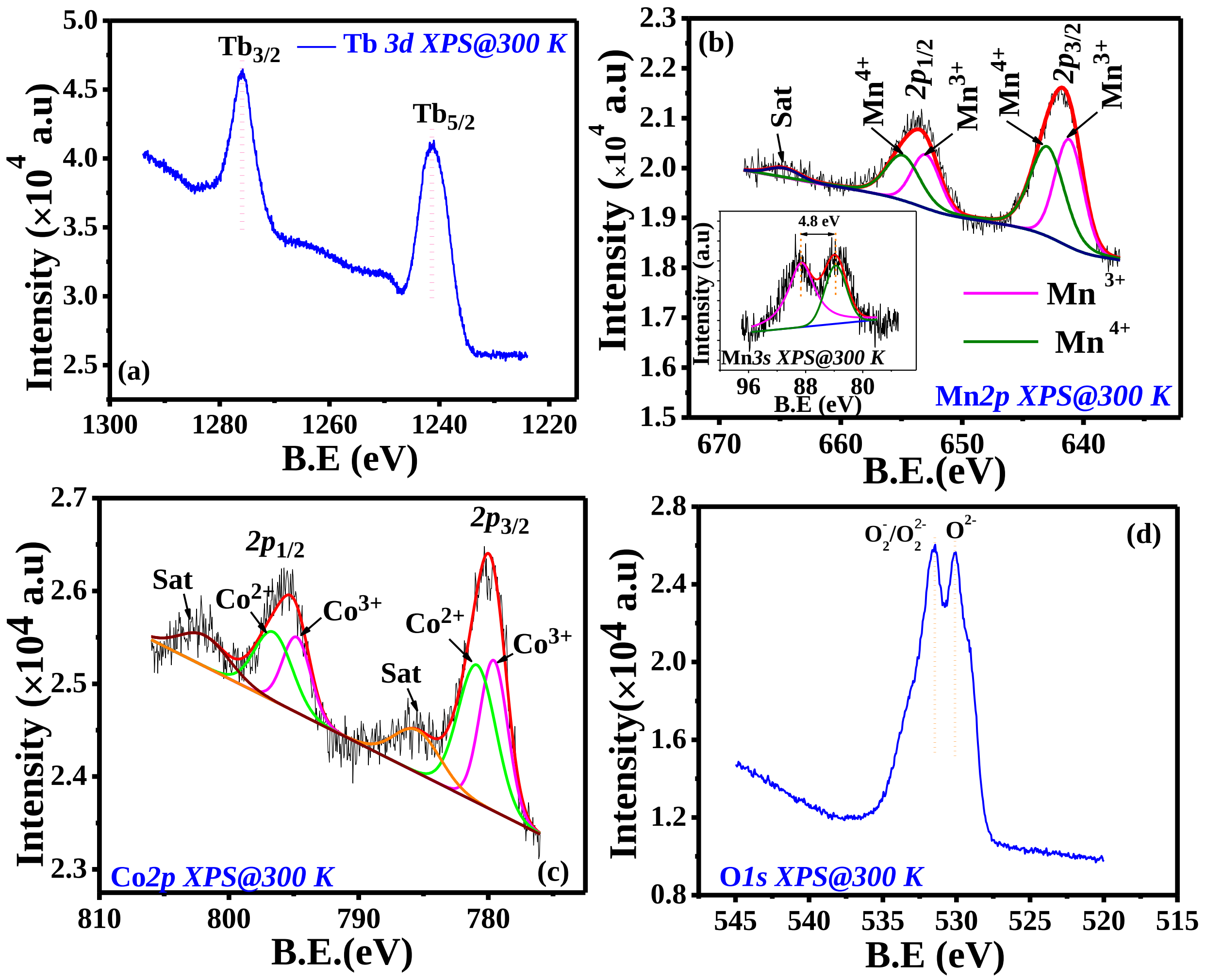

**Figure 3**

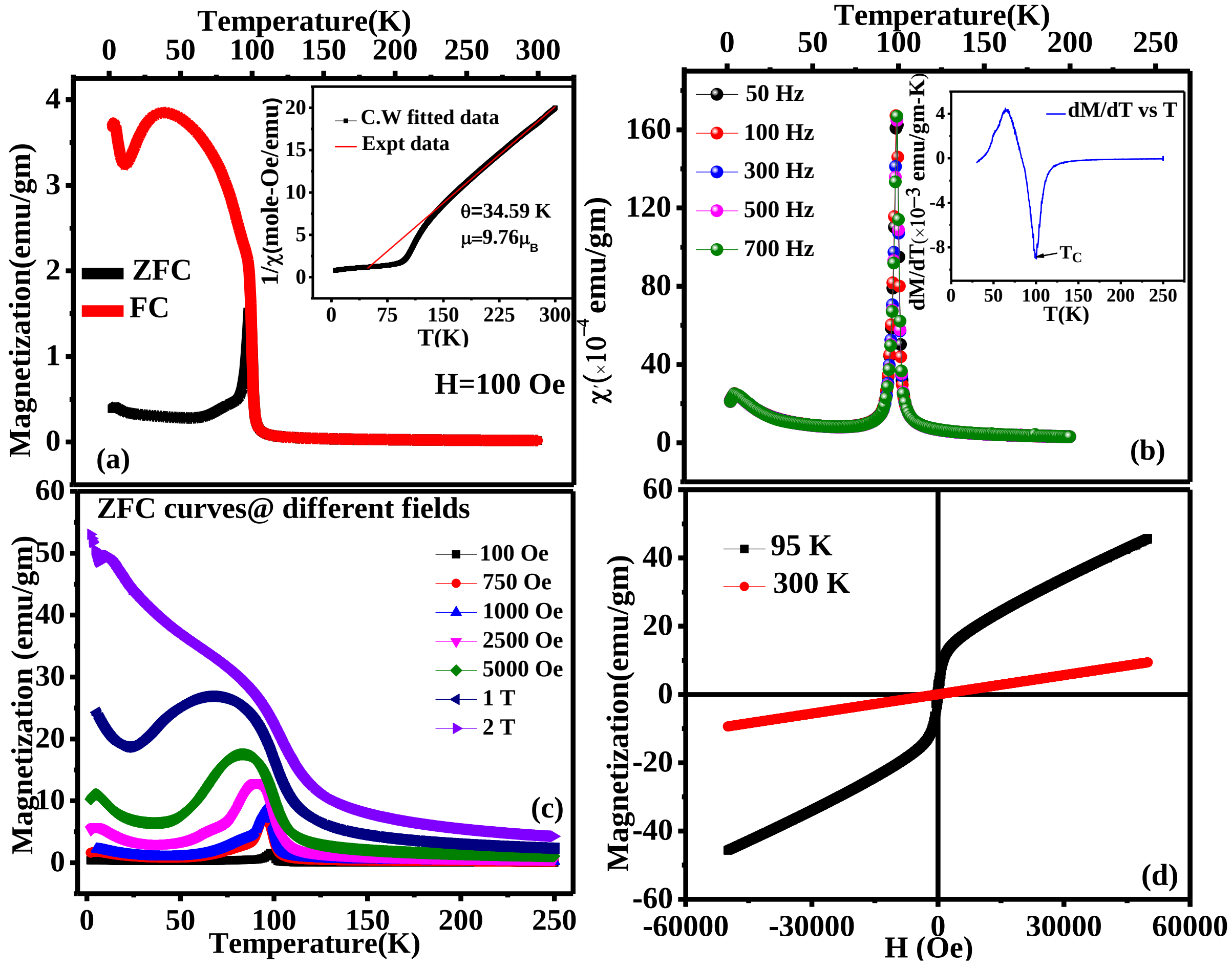

**Figure 4**

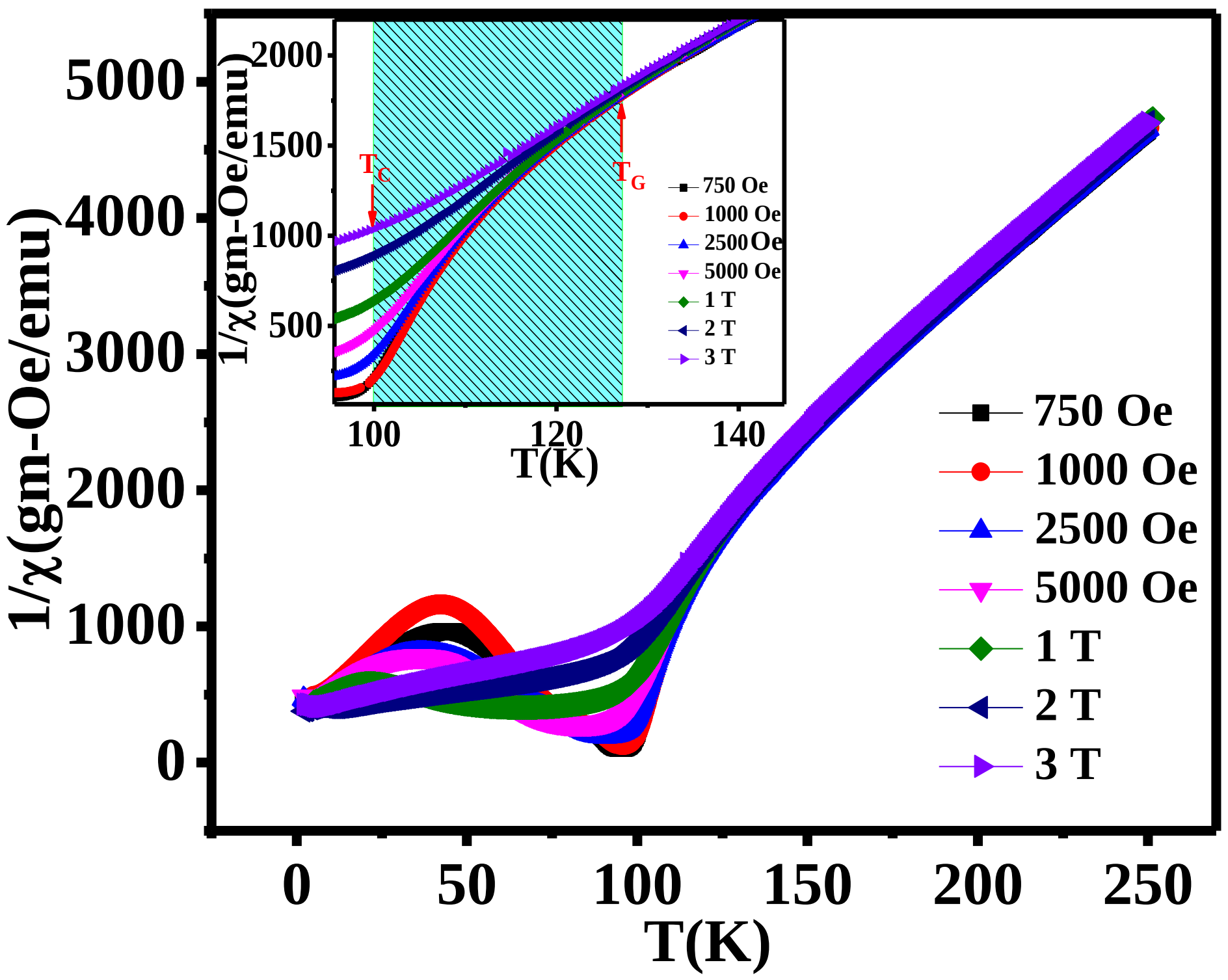

Figure 5

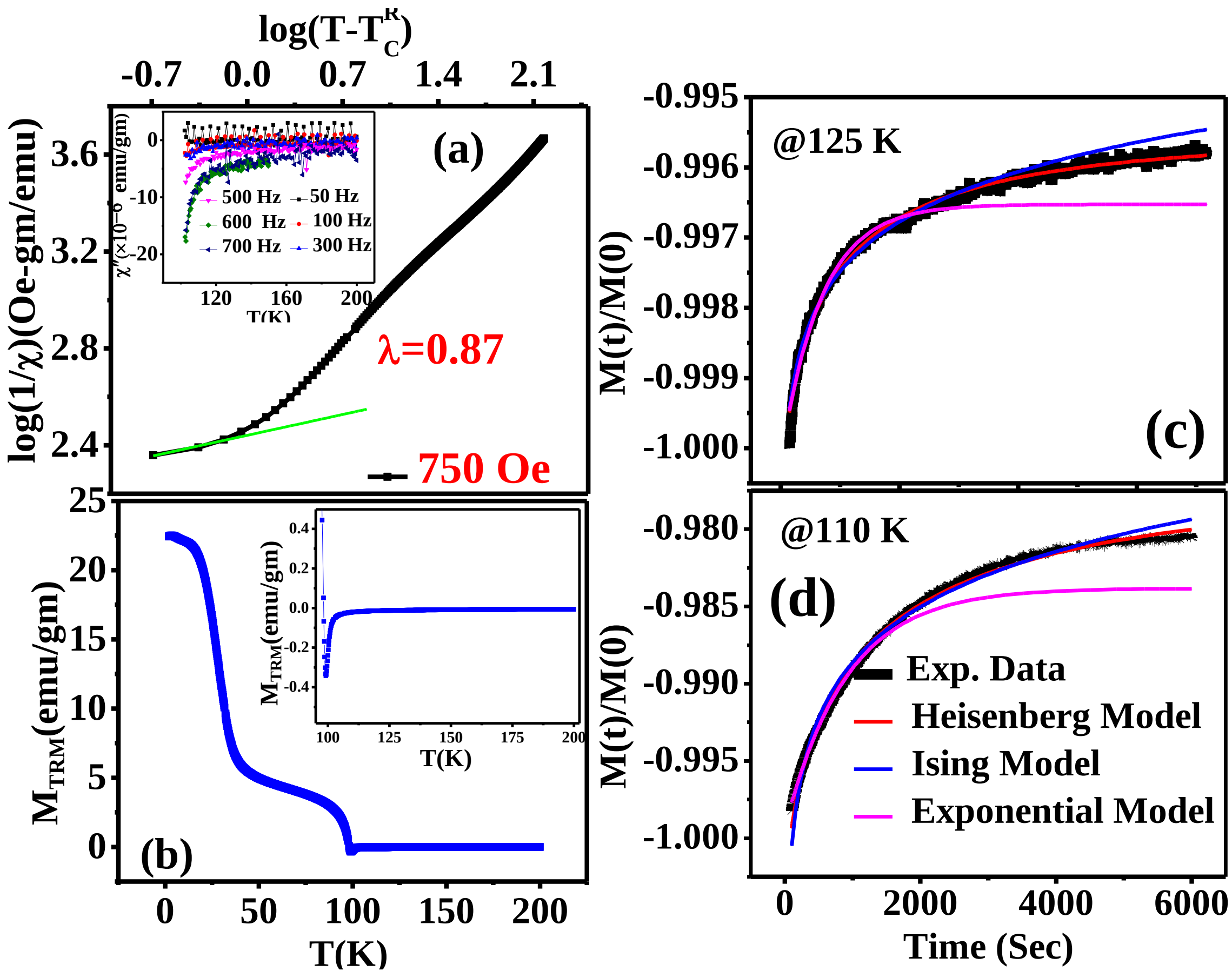

Figure 6

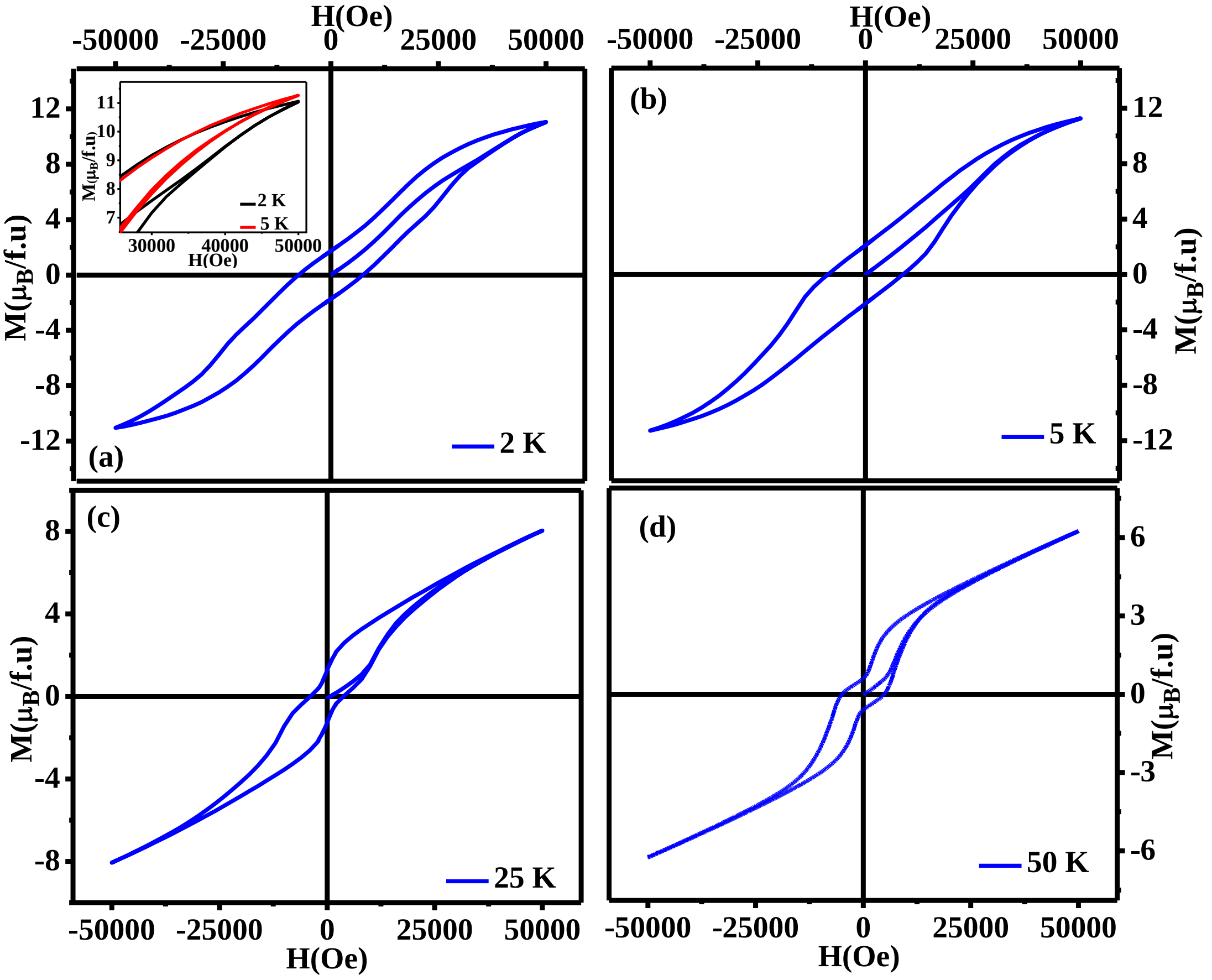

Figure 7

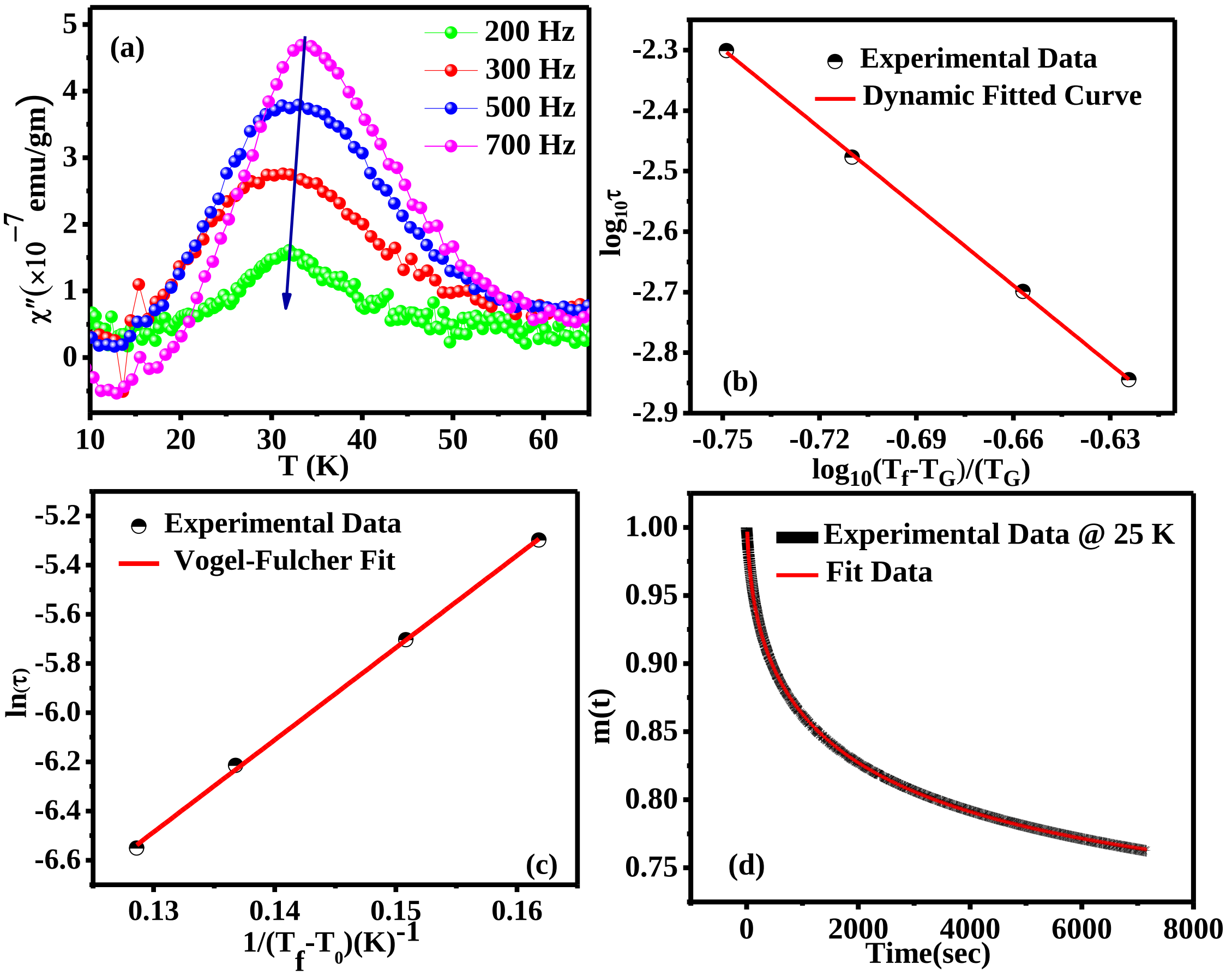

Figure 8

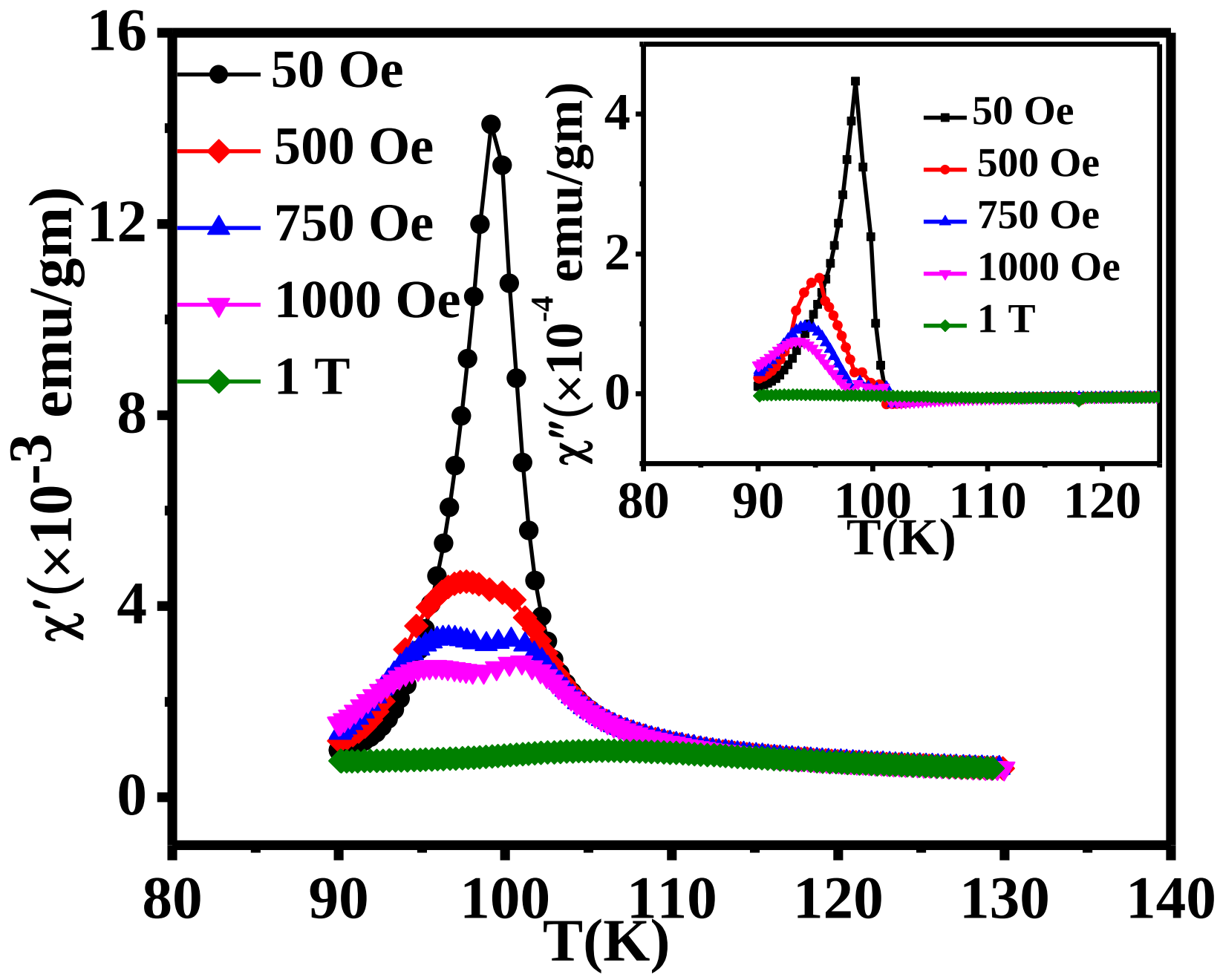

Figure 9

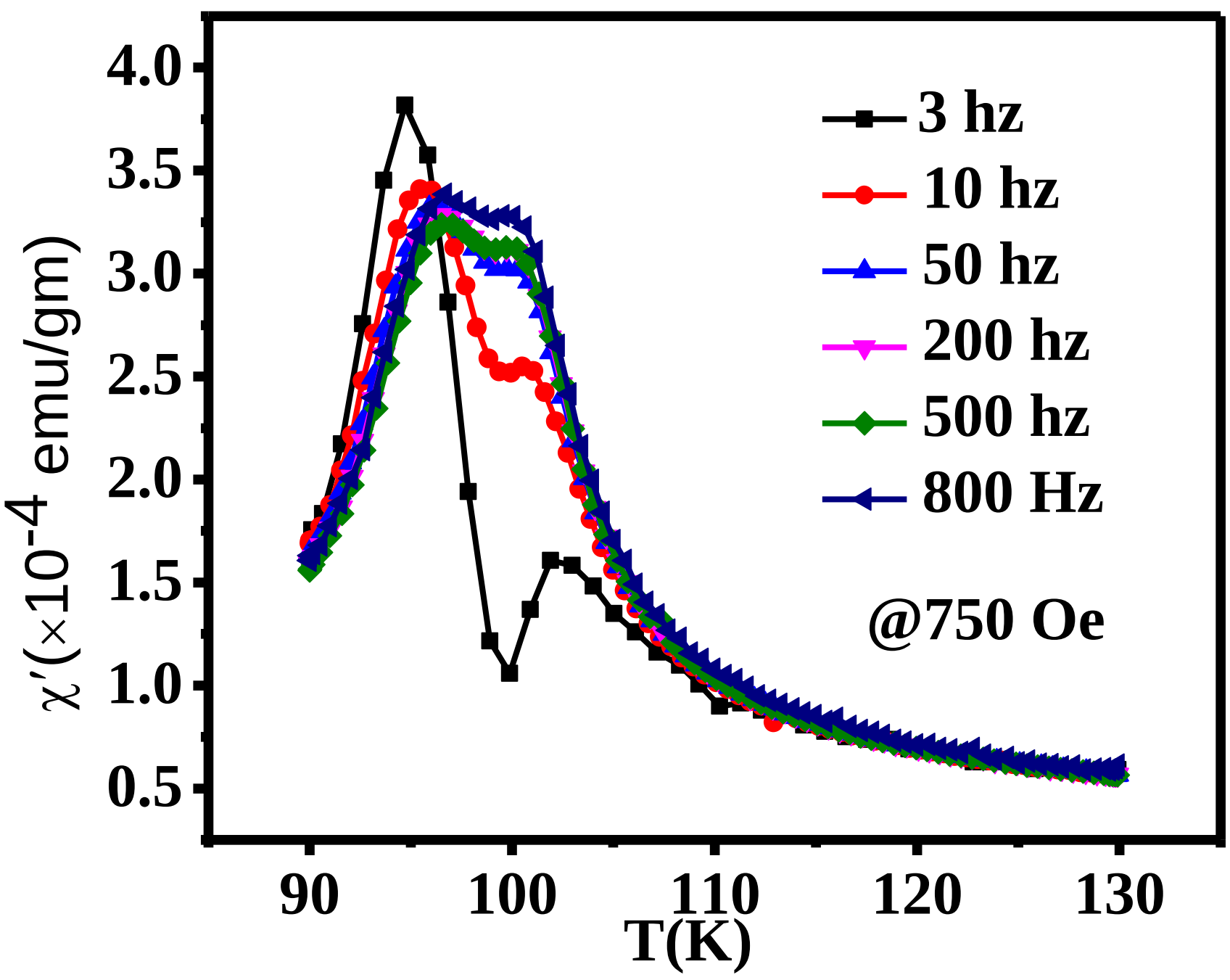

**Figure 10**